\documentstyle[aps,manuscript]{revtex}

\newcommand{\bet}{\beta}
\newcommand{\lam}{\lambda}
\newcommand{\beq}{\begin{equation}}
\newcommand{\eeq}{\end{equation}}
\newcommand{\ba}{\begin{array}}
\newcommand{\ea}{\end{array}}
\newcommand{\beqa}{\begin{eqnarray}}
\newcommand{\eeqa}{\end{eqnarray}}
\newcommand{\bd}[1]{ \mbox{\boldmath $#1$}  }
\newcommand{\wti}{\widetilde }

\begin{document}
\title{Nuclear Deformation Effects in the Cluster Radioactivity
\thanks{Send any remarks to {\tt misicu@theor1.ifa.ro}}}
\author{\c Serban Mi\c sicu
\address{
Department of Theoretical Physics,
NINPE-HH, Bucharest-Magurele, POB MG-6, Romania}
\and  Dan Protopopescu
\address{ 
Department of Physics, 
University of New Hampshire, Durham, NH 03824 , 
U.S.A.
\footnote{ on leave from Frank Laboratory of Neutron Physics, JINR,
141980 Dubna, Moscow Region, Russia}}}
\maketitle

\begin{abstract}
We investigate the influence of the nuclear deformation on the decay rates
of some cluster emission processes. The interaction between the 
daughter  and the cluster is given by a double folding potential 
including quadrupole
and hexadecupole deformed densities of both fragments.
The nuclear part of the nucleus-nucleus interaction is density 
dependent and at small distances a repulsive core in the potential will 
occur. In the frame of the WKB-approximation the assault frequency of 
the cluster will depend on the geometric properties of the potential pocket 
whereas the penetrability will be sensitive to changes in the barrier 
location.
The results obtained in this paper point out that various combinations of
cluster and daughter deformations may account for the measured values of  
the decay rate. The decay rather are however more sensitive to the 
changes in the daughter deformation due to the large mass asymmetry of 
the process.
\end{abstract}
\vskip 1.2truecm

PACS{ : 21.60.Gx , 23.70.+j , 27.90.+b }   

\section{Introduction}

The theoretical study of the heavy-cluster emission and the super-asymmetric 
fission started at the end of seventies \cite{spg80}. 
Since the beginning, this phenomenon was recognized to be a consequence of the 
shell closure of one or both fragments because of its cold nature, i.e.
the low excitation energy involved in the process. Later on, Rose and 
Jones
confirmed experimentally the existence of this new phenomenon \cite{rj84}.  
Since then many theoretical and experimental studies have been carried
out (for a review see \cite{gupgr94}). Recently it was 
advocated that the 
cluster radioactivity is not an isolated phenomenon, and must be related
to other processes like the cold fusion or cold fission\cite{cold_pred},
where the closed shell effects play a dominant role. A still opened 
problem in the study of the cluster radioactivity is represented 
by the question of the existence of only the spherical or both the 
spherical and the deformed closed shells. Although both daughter and 
emitted cluster have in many cases, at least for even-even nuclei, 
a vanishing quadrupole deformation deformation in the ground state, 
higher multipole deformations are not rulled out according to 
liquid drop calculations\cite{Mol95}. 
Until now there are no experimental data available for deformed 
daughters. 

The influence of the ground-state deformations on the fragment emission 
probability have been considered in the past\cite{shi87}. 
However, the inclusion of the deformation did not destroyed the relative
satisfactory agreement already obtained between the experimental results and 
the results of simple models without deformation. 
The first theoretical study of the cluster deformation effects 
on the WKB penetrabilities have been carried out by S\u andulescu
{\em et al.}\cite{deform} using the double folded Michigan-3 Yukawa (M3Y)
potential for a spherical daughter and a quadrupole deformed
emitted cluster. The barrier assault frequency $\nu_0$ was depending on
the deformation only by means of the first turning point location, such 
that for a deformed configuration the barrier was shifted at a larger 
interfragment distance and $\nu_0$ was decreasing.

In this paper we extend the study of the deformation effects in cluster
radioactivity by accounting also for the deformation of the daughter 
nucleus and including higher multipole deformations, like the 
hexadecupole one. The interaction between the daughter nucleus and the
cluster, in the region of small overlap and throughout the barrier  is
computed by means of a double folding potential. The nuclear part
includes a repulsive core at small distances. In this way our deformed
cluster approach supposes a cluster already formed in the potential 
pocket coming from the interplay between the Coulomb and the repulsive 
nuclear core on one hand and the attractive nuclear force on the other
hand. The depth and the width of this pocket will determine the 
assault frequency of the cluster on the barrier, through which it 
will eventually tunelate. In its turn, the penetrability will depend on 
the height and width of the barrier. Since all these geometrical 
characteristics depend sensitively on the shape of the fragments we will 
investigate in this paper the modification induced by the quadrupole and
hexadecupole deformations of the fragments on the pocket and the barrier
and finally compute decay rates for several cluster emitters.

\section{Cluster-Daughter Double-Folding Potential }

The nuclear interaction between the daughter and the cluster
can be calculated as the double folding integral of ground state 
one-body densities  $\rho_{1(2)}(\bd{r})$ of heavy ions
as follows
\beq
U_N(\bd{R}) = \int d{\bd r}_{1} \int d{\bd r}_{2}~
\rho_{1} ({\bd r}_{1}) \rho_{2} ({\bd r}_{2}) v({\bd s}) 
\eeq
where $v$ is the $NN$ effective interaction and the separation distance
between two interacting nucleons is denoted by 
${\bd s}={\bd r}_{1}+{\bd R}-{\bd r}_{2}$, $R$ being the
centre-to-centre distance. In the past a $G$-matrix M3Y effective interaction 
was used to discuss light and heavy cluster radioactivity\cite{deform,m3y}. 
This interaction contains isoscalar and isovector Yukawa functions in each
spin-isospin $(S,T)$ channel and an exchange component coming from the 
one-nucleon knock-on exchange term. However, this interaction is based
on density-independent nucleon-nucleon forces and consequently 
very deep nucleus-nucleus potentials are obtained. 
A double folding potential based on the effective Skyrme interaction will
contain a repulsive core which would prevent, according to the Pauli
principle, a large overlap of the two interacting nuclei\cite{Adam96}. 
Thus the nuclear potential between two ions contains
an attractive part and a repulsive one. Neglecting the spin
dependence, it can be written as 
\beq
U_{N}(\bd{R}) = C_0\left\{ \frac{F_{in}-F_{ex}}{\rho_{00}}
\left ((\rho_{1}^2*\rho_2)(\bd{R}) +
       (\rho_{1}*\rho_2^2)(\bd{R})\right)\\
 + F_{ex}(\rho_{1}*\rho_2)(\bd{R})\right\}
\eeq
where $*$ denotes the convolution of two functions $f$ and $g$, i.e.
$(f*g)(\bd{x})=\int f(\bd{x}')g(\bd{x}-\bd{x}')d\bd{x}'$.
The constant $C_0$ and the dimensionless parameters $F_{in},F_{ex}$ are
given in ref.\cite{Adam96}. To solve this integral we consider the inverse 
Fourier transform
\beq
U_{N}(\bd{R})=\int e^{-i\bd{q}\cdot\bd{R}}{\wti U}_N(\bd{q})d\bd{q}
\eeq
where the Fourier transform of the local Skyrme potential ${\wti
U_N}(\bd{q})$ can be casted in the form
\beq
{\wti U_{N}}(\bd{q})=C_0\left\{\frac{F_{in}-F_{ex}}{\rho_{00}}
\left({\wti {\rho_{1}^2}}(\bd{q}){\wti \rho_2(-\bd{q}}) + 
{\wti \rho_{1}}(\bd{q}){\wti {\rho_2^2}}(-\bd{q})\right )
+F_{ex}{\wti \rho_{1}}(\bd{q}){\wti \rho_2}(\bd{q})\right \}
\eeq 
Here ${\wti \rho}(\bd{q})$ and ${\wti {\rho^2}}(\bd{q})$ are Fourier
transforms of the nucleon densities $\rho(\bd{r})$ and squared nuclear 
densities $\rho^2(\bd{r})$. Expanding the nucleon densities for
axial-symmetric distributions in spherical harmonics we get 
\beq
\rho(\bd{r})=\sum_{\lam}\rho_{\lam}(r)Y_{\lam 0}(\theta,0)
\eeq 
Then
\beqa
{\wti \rho}(\bd{q}) & = &4\pi\sum_{\lam}i^{\lam}Y_{\lam 0}(\theta_q,0)
\int_0^\infty r^2 dr \rho_{\lam}(r) j_{\lam}(qr)
\\
{\wti {\rho^2}}(\bd{q})& = & \sqrt{4\pi}\sum_{\lam}
\frac{i^{\lam}}{\hat \lam}
Y_{\lam 0}(\theta_q,0) \sum_{\lam'\lam''}
{\hat \lam'}{\hat \lam''}(C_{0~0~0}^{\lam\lam'\lam''})^2\times
\nonumber\\
&&\int_0^\infty r^2 dr \rho_{\lam'}(r)\rho_{\lam''}(r) j_{\lam}(qr)
\eeqa 

In this paper we take the one-body densities for both daughter and cluster
as two-parameter Fermi distributions in the intrinsic frame for axial
symmetric nuclei
\beq
\rho(\bd{r})=\frac{\rho_{00}}{1+\exp((r-R(\theta))/a)}
\eeq
Here $\rho_{00} = 0.17$ fm, $a$ denotes the diffusivity which is taken
to be 0.63 for the daughter and 0.67 for the cluster, and  
\beq
R(\theta)=R_0\left ( 1+\bet_2\sqrt{\frac{5}{4\pi}}P_2(\cos\theta)+
\bet_4\frac{3}{\sqrt{4\pi}}P_4(\cos\theta)\right )
\eeq
is the parameterization of the nuclear shape in quadrupole $\bet_2$ 
and hexadecupole $\bet_4$ deformations. Here $R_0=r_0A^{1/3}$ with
$r_0$ computed by means of a liquid drop prescription\cite{Mol95}.

\section{Calculus of decay constants}

We adopt a modified Gammow approach\cite{gupgr94}  which is based on the
idea that the cluster is pre-born, with a certain probability
$P_0$, in the pocket of the Skyrme+Coulomb potential and later on it
tunnels through an essentially one-dimensional barrier. 
Although in the present case, having deformed fragments, we deal 
with a multi-dimensional penetration problem, we can disregard the 
orientation effects in a first approximation. 
It is not difficult to show that the value of the penetrability  
is maximized on the fission path corresponding to daughter and 
cluster having their symmetry axes oriented along the inter-fragment
axis. For such a configuration the top of the barrier will attain a 
minimum and its location is found at a larger interfragment distance.

Consequently the decay rate $\lam$ will be defined as follows:
\beq
\lam=\nu_0P_0P   
\label{lambda}
\eeq
where $\nu_0$ is the assault frequency with which the cluster
bombards the walls of the potential pocket. It is given by the inverse 
of the classical period of motion 
\beq
T_{0} =\int\limits_{r_{t_1}}^{r_{t_2}}
dr\sqrt{\frac{2\mu}{Q-U(r)}}
\label{niu0}
\eeq
where $\mu$ is the reduced mass of the cluster-daughter pair and 
$r_{t_1}$ and $r_{t_2}$ are the inner turning points, where the
potential curve intersects the $Q$-value (see Fig.1). 
Thus, in our model, $\nu_0$ depends sensitively  on the size of the
potential pocket. The barrier penetrability is computed in the frame
of the WKB approach
\beq
P = \exp\left( -2\int\limits_{r_{t_2}}^{r_{t_3}}d r
\sqrt{\frac{2\mu}{\hbar^2}(U(r)-Q)} \right )
\eeq
where $r_{t_3}$ is the outer turning point.

The calculus of the preformation probability $P_0$ is usually based
on elaborated microscopic models. Since its calculation is beyond the
purpose of this paper, we limit ourselves to a simple empirical formula 
proposed by Blendowske $et. al.$ \cite{bf91} for $\alpha,{}^{12}\mbox{C}, 
{}^{14}\mbox{C} $ and ${}^{16}\mbox{O}$ clusters
\beq
P_0=(P_0^\alpha)^\frac{A_{c}-1}{3}~~~~~~ (A_{c}\leq 28)
\eeq
where the subscript $c$ refers to the cluster and the
$\alpha$-spectroscopic
factor is estimated as
\beq(P_0^\alpha)^{even}=6.3\times 10^{-3}~~~~~~~\mbox{and}
~~~~~~~(P_0^\alpha)^{odd}=3.2\times 10^{-3}
\eeq
In what follows we consider the $^{14}\mbox{C}$-decay of $^{224}\mbox{Ra}$.
 
In Fig.1 we plotted a family of potential curves $U(R)$
for several quadrupole deformations $\beta_2^C$ of the cluster
$^{14}\mbox{C}$ and a fixed hexadecupole deformation of the daughter,
chosen to be $\beta_4^D=0.008$, i.e. the ground state value 
for $^{210}\mbox{Pb}$.
As one can see on this plot, the increase of $\beta_2^C$ from
negative to positive values is accompanied by the lowering of the
barrier, while the bottom of the pocket goes down further.
The decay rate is influenced mainly by the changes with deformation 
in the region between the first turning point and the top of the 
barrier. 

For comparison, the next plot, Fig.2, shows the variation
of the interaction potential with the hexadecupole deformation of the cluster.
The variation brought by the hexadecupole deformation is slightly different.  
Like in the previous case, the barrier lowers with $\beta_4^C$,
but the bottom of the pocket rises. Obviously, this will affect
the values of life-times in a different way than the one $\beta_2^C$ does. 
The assault frequency $\nu_0$ will decrease with increasing cluster hexadecupole
deformation, but not enough to compensate the increase of penetrability.
In overall the decay rate will increase with $\beta_4^C$ slower than
with $\beta_2^C$.

Plots of the interaction potential $U(R)$ for different pairs
($\beta_2^{D}$;$\beta_2^{C}$) of quadrupole deformation are shown in Fig.3.
One can compare the magnitude of the effect of both cluster and daughter
quadrupole deformations and the changes which occur when we pass
from prolate to oblate deformation.
One can notice that the quadrupole deformation $\beta_2^C$  of the cluster
acts mainly on the right wall of the pocket, while the modification of the  
quadrupole deformation of the daughter nucleus push in opposite
directions both walls of the pocket.

The difference between positive and negative quadrupole and hexadecupole
deformations of the daughter nucleus can be easily understood from
Fig.4.
We observe different type of modifications in the shape of the pocket,
corresponding to $\beta_2$ and $\beta_4$ deformations, respectively.
The quadrupole and hexadecupole deformations of the daughter  change 
the depth of the potential pocket in the same manner, in comparison
with the case of cluster deformations which, as we saw in figures 
1 and 2, move the bottom of the pocket in opposite directions.

A plot of calculated $\lambda$, as a function of the hexadecupole 
deformation of the daughter nucleus $\beta_4^D$, for several values of
$\beta_2^D$ and a spherical cluster, is drawn in logarithmic scale in Fig.5. 
One can see, for example, that if the daughter nucleus has no quadrupole
deformation a hexadecupole deformation $\beta_4^D\approx0.04$ may account 
for the experimental decay rate. Values of the decay rate 
close to the experimental one  can be attained also by
combinations of non-vanishing quadrupole and hexadecupole deformations.

The next figure presents the variation of the calculated decay rate $\lambda$ with
cluster hexadecupole deformation for four fixed values of $\beta_2^{C}$(see Fig.6). 
This time the daughter nucleus is taken to be spherical. 
Recalling the observations made earlier, in connection with figures 1 and 2,
on the modification of the potential due to the quadrupole and hexadecupole
deformations one may understood why $\lambda$ increases faster with 
$\beta_4^{D}$ than with $\beta_4^{C}$. Therefore one may 
infer that it is less favorable to emit the cluster with a non-zero 
hexadecupole deformation. 
 
In Table I we selected some of the most favorable cases
for the calculated decay rates for several cluster decay reactions. 
As we expected, prolate deformations favor the decay. 
In the case of $^{14}$C decay of $^{224}$Ra deformations between -0.04 and 
0.04 in $\beta_2^D$ give us decay rates which match the experimental one
provided $\beta_4^D$ is taken to be not to large compared to the ground 
state value, i.e. 0.008. We may also consider a case when the cluster 
is deformed and the daughter is spherical.
In the case of $^{14}$C decay of $^{226}$Ra, a hexadecupole deformation
not too large than its ground state value still accounts for the experimental
value although all other deformations are zero.

The fact that the decay rates are more sensitive to the daughter 
deformations can be easily understood on ground of the mass asymmetry
of the decaying system. An easy to follow explanation of this fact 
is given if we consider only the Coulomb part of the barrier. The 
asymptotic part of the barrier, including the region where the third 
turning point $r_{t_3}$ is located, is determined essentially by the Coulomb
multipoles. Introducing the quadrupole moments of the charge density
\beq
Q_{2}^{D,C}=\sqrt{\frac{4\pi}{5}}
\int_{0}^{\infty} r^{2} dr \rho_{2}^{D,C}(r) r^{2}
\eeq
the potential will behave for $R\rightarrow\infty$ as
\beq
(C_{000}^{202})^{2}\frac{Z_{C}Q_{2}^{D}+Z_{D}Q_{2}^{C}}{R^3}.
\eeq
Since the quadrupole moments of the daughter $Q_2^{D}$ and cluster
$Q_2^{C}$ depend on their quadrupole deformations $\beta_2$ it is 
obvious that due to the asymmetry in mass and charge 
($Z_{C}/Z_{D}\approx 0.1$), the deformation of the daughter will have
a larger influence on the barrier and eventually on the value of the decay 
rate.

 \section{Conclusions}

The aim of this paper was to extend previous studies of deformation
effects in cluster radioactivity by considering also the deformation 
of the daughter nucleus and to include the next higher even deformation,
the hexadecupole one. Considering that the cluster is pre-born in the 
potential pocket produced by the interplay between repulsive and attractive 
forces, we investigated the modifications induced by deformations  
on the specific potential that we employ. 
The computed decay rates depend on the assault frequency, which varies with 
the pocket depth, and on the penetrability, which changes with the barrier 
height. 
We showed that the experimental values can be reproduced for several 
selections of the deformations. If we maintain the cluster spherical
and vary the quadrupole and/or hexadecupole deformations of the daughter
nucleus we may reach the experimental value within a reasonable
range of deformations parameters. Conversely, if the daughter is 
spherical, then only for significant values of $\beta_2^{C}$ we can
reproduce the experimental value without taking into account $\beta_4^{C}$. 
One might conclude that deformed states of the daughter are
needed to be considered in order to reach the
experimental $\lambda$ within reasonable values of  
cluster deformation. Configurations with a spherical cluster and  
only a hexadecupole deformation for the daughter are likely to occur. 
Also combinations of small deformations of both fragments should not be
discarded.
In conclusion, the study carried out in this paper points 
mainly to the importance of the daughter nucleus deformations, 
and especially its hexadecupole one.  

\section{Acknowledgments} One of the authors (\c S.M.) would like to
acknowledge the hospitality of Prof.R.Jolos at the Bogoliubov
Laboratory for Theoretical Physics during the completion of this work 
and to express its gratitude to Dr.A.Nasirov for many enlightening discussions.
We are also very indebted to Dr.F.Carstoiu who provided us the routines 
for calculating the Coulomb part of the interaction potential.

\begin{table}
\caption{\footnotesize Deformations of clusters and daughters which fit
the calculated decay rate (see eq.(\ref{lambda})) to the experimental one 
$\lambda_{exp}$. Empty spaces in the table mean
null values.}
\label{tabel}
\vspace{5mm}
\begin{tabular}{|ccccccc|}
\hline
Decay & \multicolumn{2}{c}{Cluster}& &\multicolumn{2}{c}{Daughter}&
$\lambda_{exp}$\\   
\cline{2-3}\cline{5-6}
&$\beta_2^C$&$\beta_4^C$&&$\beta_2^D$&$\beta_4^D$&$(s^{-1})$\\
\hline
&&&&&&\\
$^{224}$Ra$\rightarrow^{14}$C+$^{210}$Pb& 0.160 & 0.056 &&  & & 
9.50$\cdot$10$^{-17}$\\
       & & &&  -0.040 & 0.072 &\\
       & & &&   & 0.040 &\\
       & & &&   0.040 & 0.007 &\\
$^{226}$Ra$\rightarrow^{14}$C+$^{212}$Pb& & & & & 0.029 
&3.50$\cdot$10$^{-22}$\\
$^{234}$U$\rightarrow^{24}$Ne+$^{210}$Pb&-0.215 & 0.155 && & 0.098 
&5.94$\cdot$10$^{-26}$\\
&0.170&  && & 0.010 &\\
$^{234}$U$\rightarrow^{28}$Mg+$^{206}$Hg& 0.323 & -0.136 && -0.033 &  &
2.00$\cdot$10$^{-26}$\\
&0.113& && 0.080 & 0.080 &\\
$^{230}$Th$\rightarrow^{24}$Ne+$^{206}$Hg&-0.215 & 0.155 && 0.123 & &
1.60$\cdot$10$^{-25}$\\
& &&& 0.063 & &\\
\hline
\end{tabular}
\end{table}

\newpage
\centerline {FIGURE CAPTIONS}
\vskip 1truecm

${\bf Fig.~1.}$ Variation of the interaction potential $U(R)$
with the quadrupole deformation of the cluster, $\beta_2^C$,
when the ground hexadecupole deformation for $^{210}$Pb is kept fixed,
i.e. $\beta_4^{D}=0.008$. The distance $R$ is measured between the centers
of mass of the two nuclei. The horizontal line at $ 30.53$ MeV is the
$Q$-value of the decay.
\vskip 1.0 cm

 ${\bf Fig.~2.}$ Variation of the interaction potential $U(R)$
with the hexadecupole deformation of the cluster, $\beta_4^C$. All
deformation parameters, except $\beta_4^D$=0.008, are zero.  
\vskip 1.0 cm

${\bf Fig.~3.}$ Plots of the interaction potential $U(R)$
for different pairs of  daughter-cluster quadrupole deformation
($\beta_2^{D}$;$\beta_2^{C}$).
\vskip 1.0 cm

 ${\bf Fig.~4.}$ Comparison between the effects of
quadrupole and hexadecupole deformations of the daughter
nuclei. Here, contrary to the case when the cluster is deformed,
the quadrupole and hexadecupole deformations modify
the depth of the pocket in the same manner.
\vskip 1.0 cm

${\bf Fig.~5.}$ 
The dependency of $\lambda$ on $\beta_4^D$ in logarithmic scale for several 
values of daughter's quadrupole deformation. The cluster is
spherical in all cases. The horizontal line represents
the experimental value for the discussed decay, 
$\lambda_{exp}=9.50\times 10^{-17}$ s$^{-1}$.
\vskip 1.0 cm

${\bf Fig.~6.}$ 
Plots of the variation of calculated decay rate $\lambda$ with
hexadecupole deformation of cluster for four fixed values
of $\beta_2^{C}$. The daughter nucleus is taken spherical.

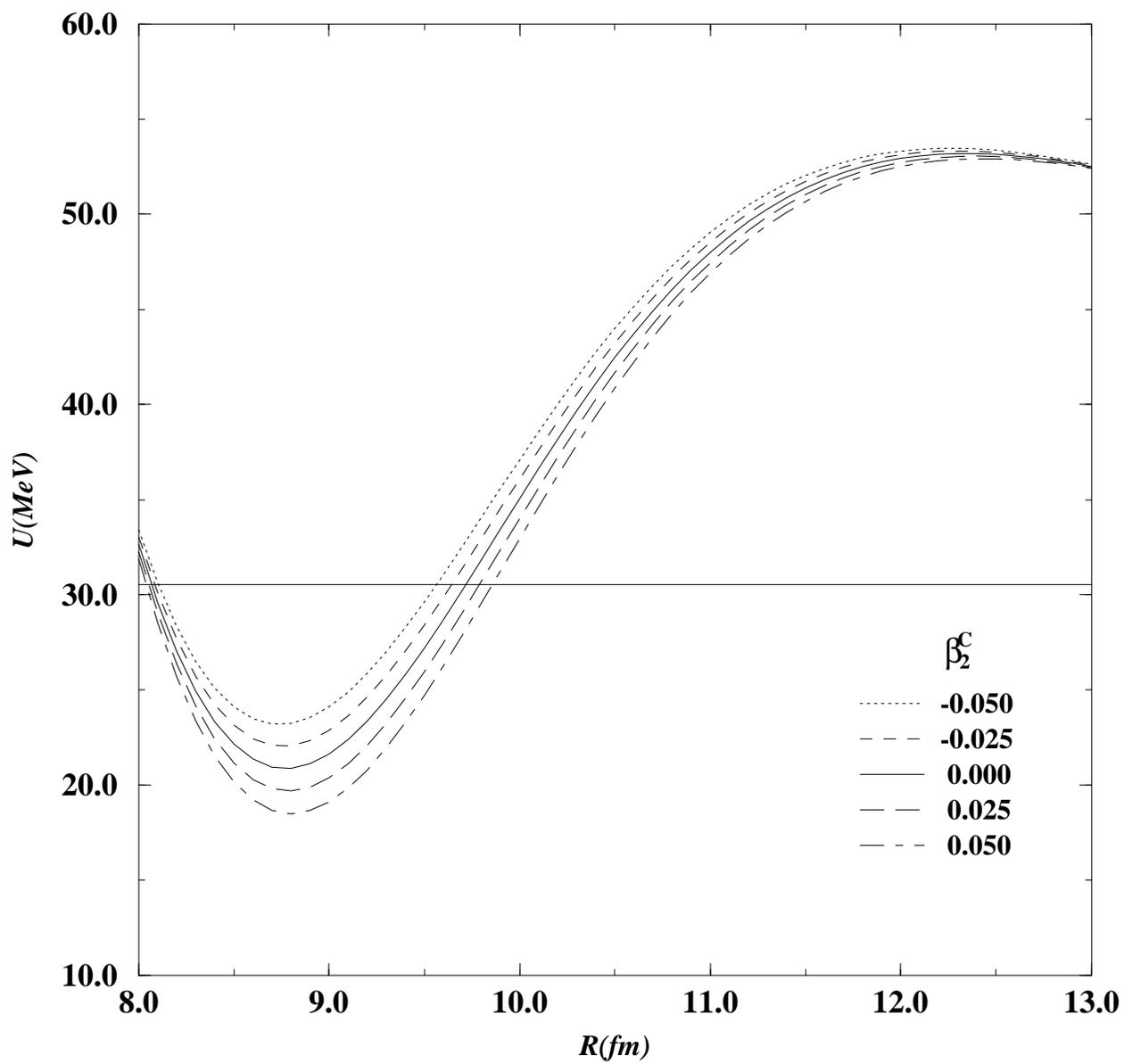
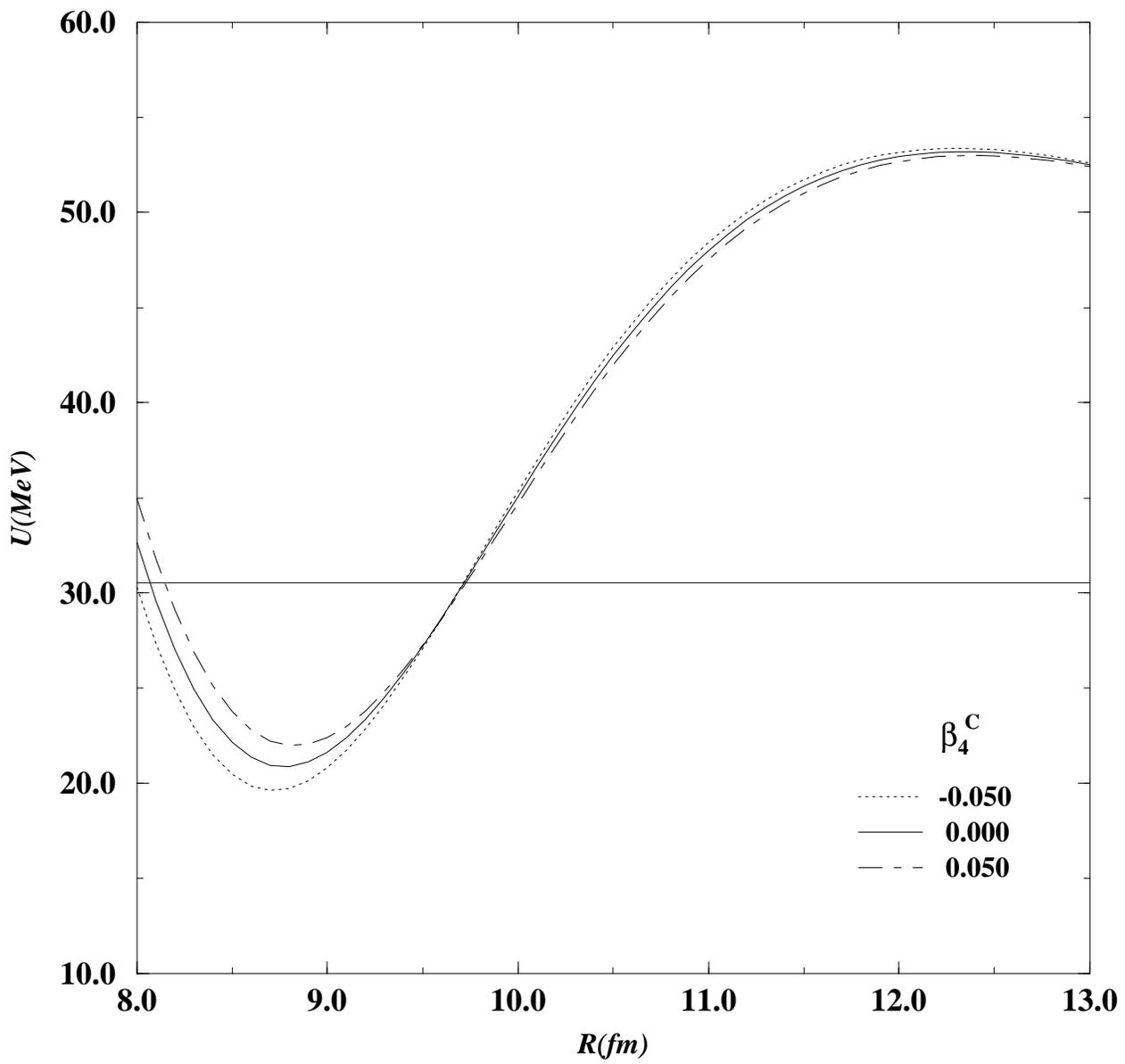
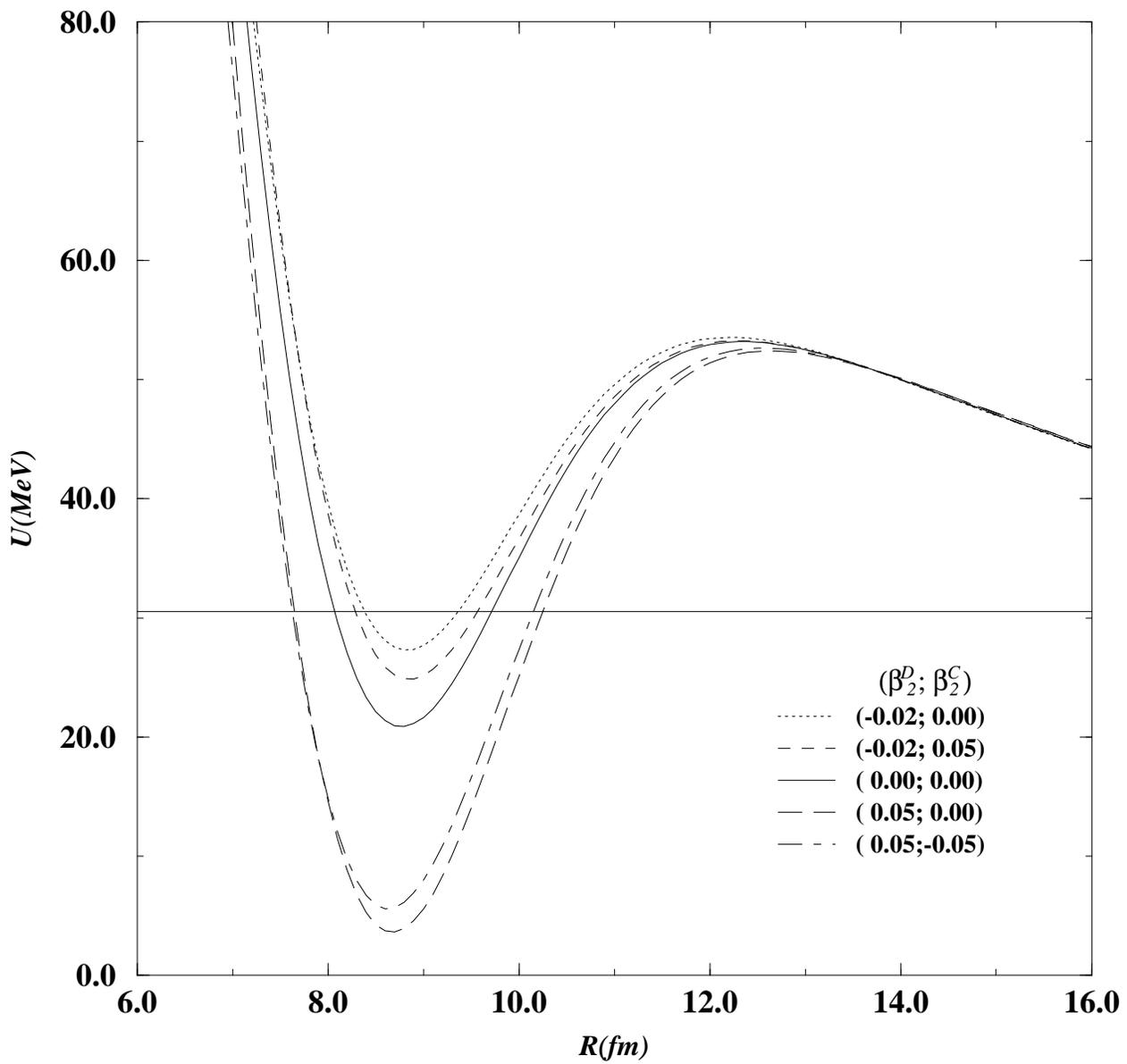
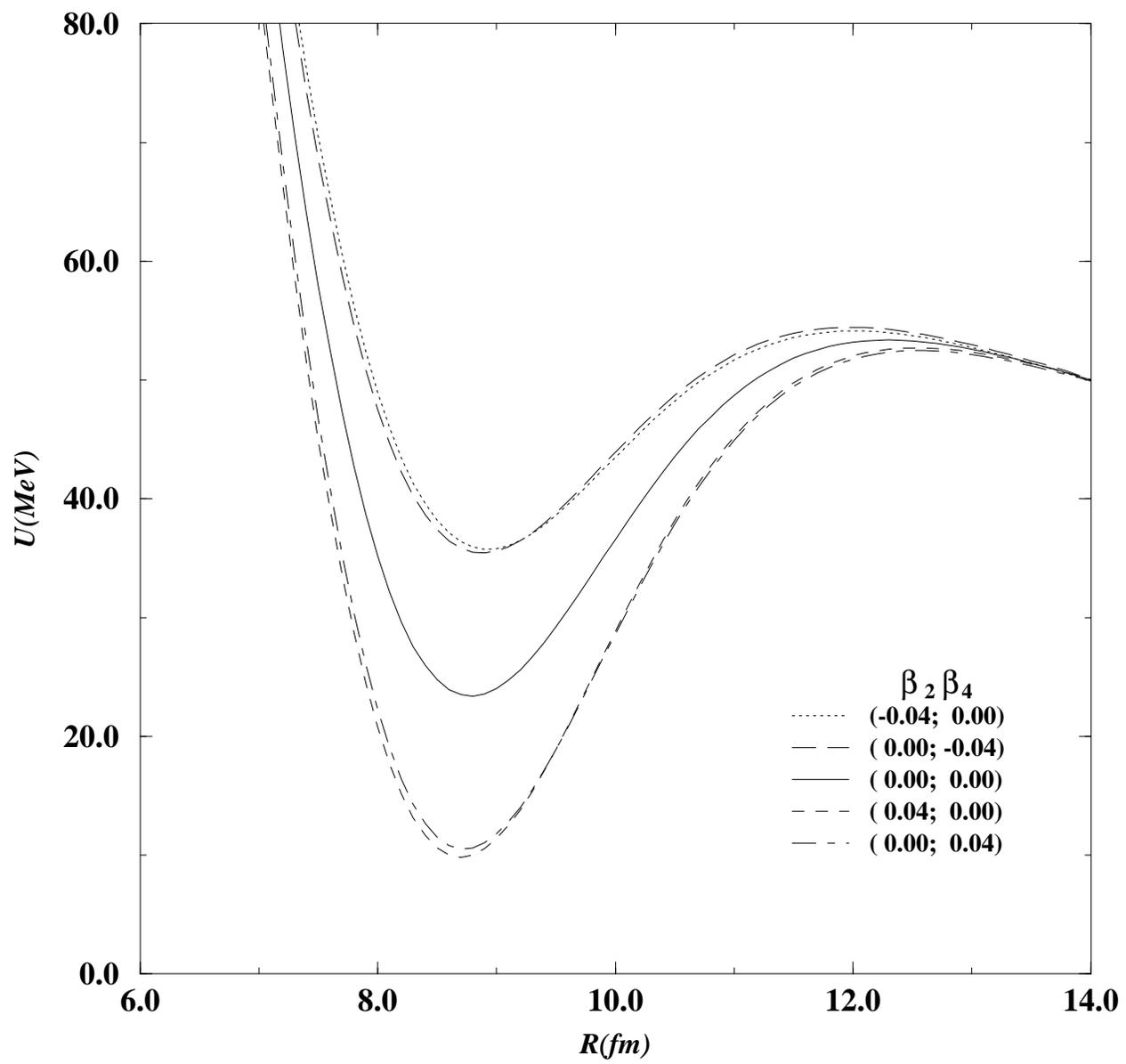
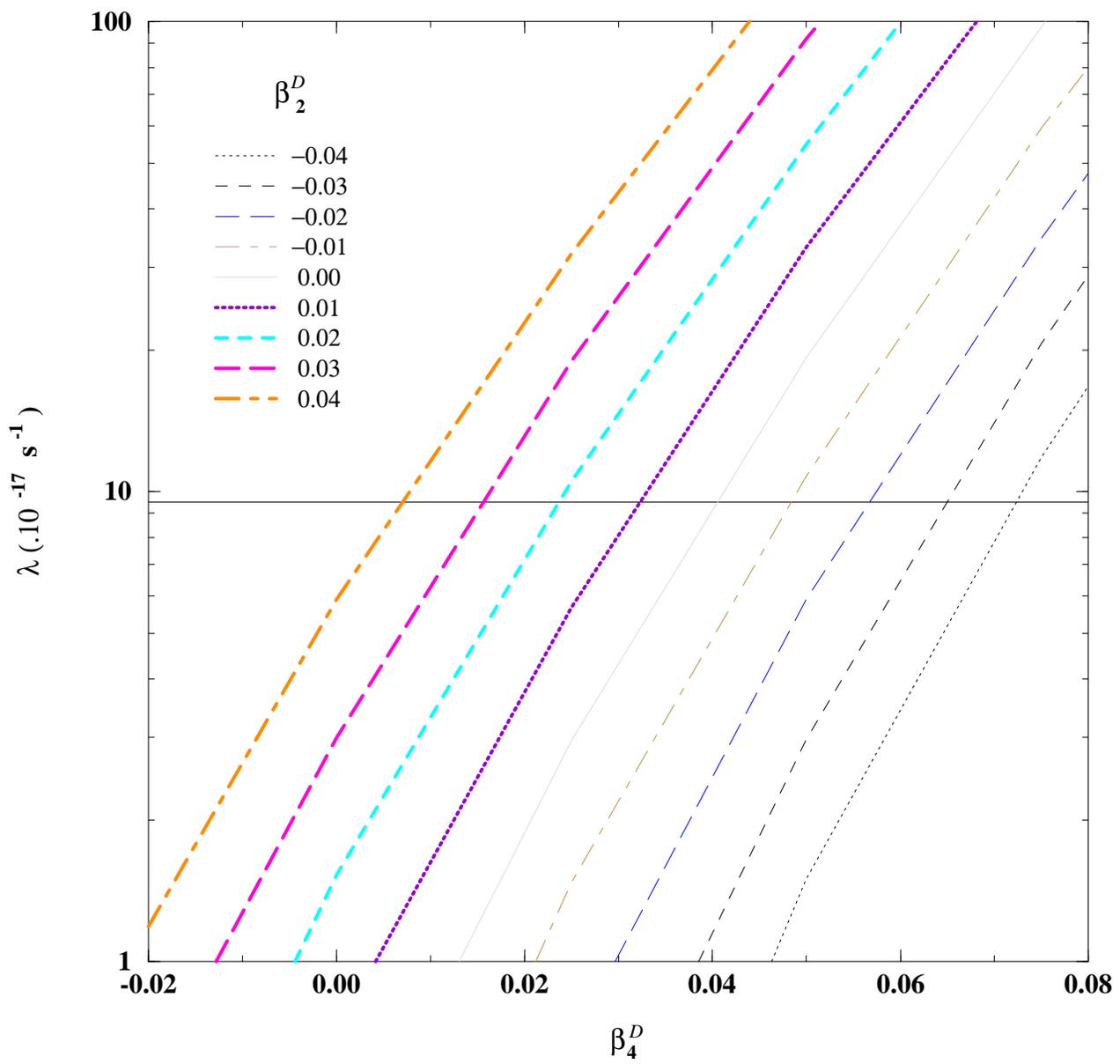
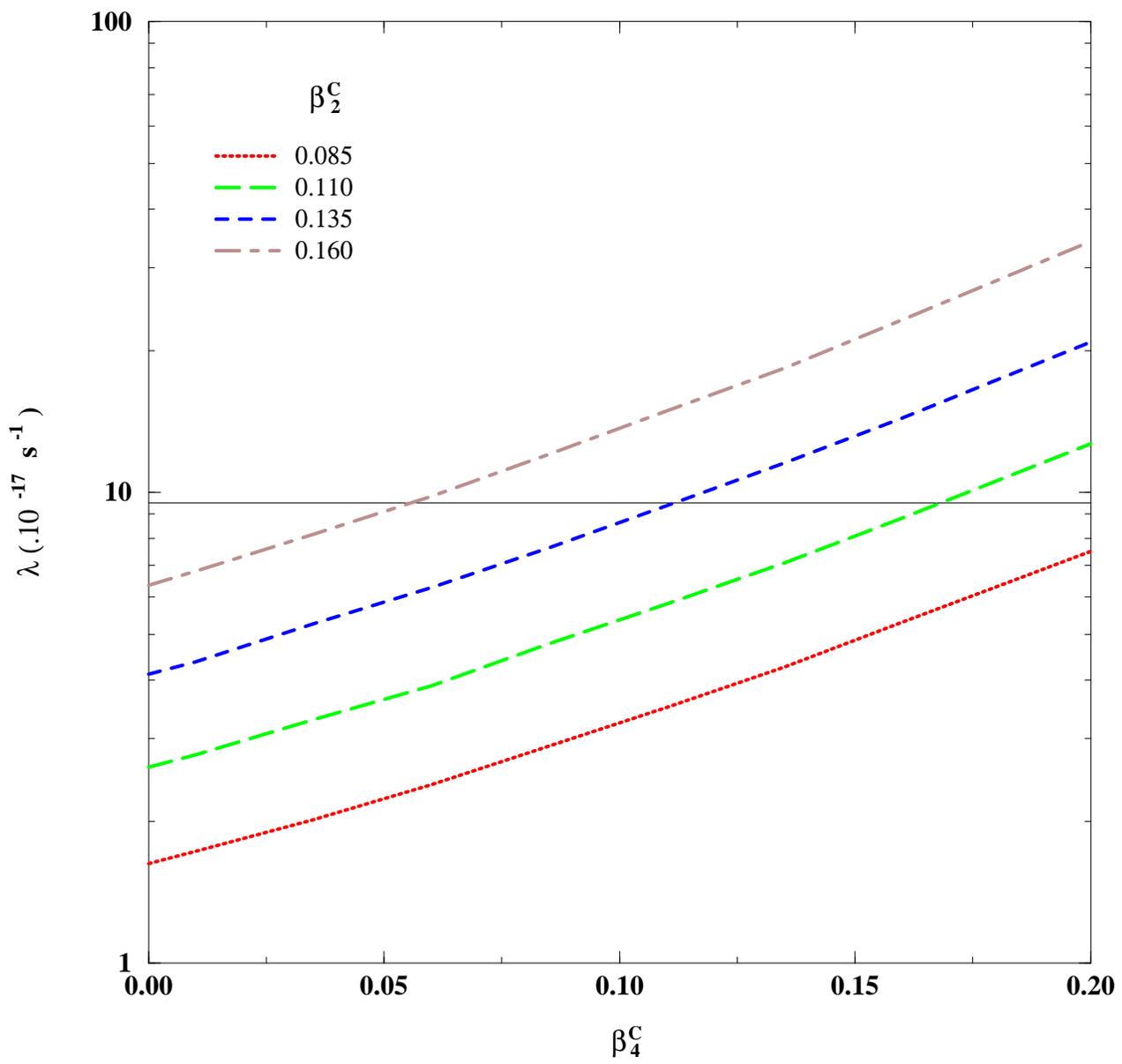
\end{document}